\documentclass[manuscript]{acmart}
\AtBeginDocument{%
  \providecommand\BibTeX{{%
    \normalfont B\kern-0.5em{\scshape i\kern-0.25em b}\kern-0.8em\TeX}}}

\copyrightyear{2021} 
\acmYear{2021} 
\setcopyright{acmlicensed}\acmConference[RecSys '21]{Fifteenth ACM Conference on Recommender Systems}{September 27-October 1, 2021}{Amsterdam, Netherlands}
\acmBooktitle{Fifteenth ACM Conference on Recommender Systems (RecSys '21), September 27-October 1, 2021, Amsterdam, Netherlands}
\acmPrice{15.00}
\acmDOI{10.1145/3460231.3474242}
\acmISBN{978-1-4503-8458-2/21/09}

\usepackage{wrapfig}

\begin{document}

\title[Page-level Optimization]{Page-level Optimization of e-Commerce Item Recommendations}

\author{Chieh Lo, Hongliang Yu, Xin Yin, Krutika Shetty, Changchen He, Kathy Hu, Justin Platz, Adam Ilardi, Sriganesh Madhvanath}
\email{{chielo, honyu, xinyin, krshetty, chhe, xiyuhu, jplatz, ailardi, smadhvanath}@ebay.com}
\affiliation{%
  \institution{eBay Inc.}
  \city{New York City}
  \state{New York}
  \country{USA}
  \postcode{10011}
}
\renewcommand{\shortauthors}{Chieh and Yu, et al.}

\begin{abstract}
  The item details page (IDP) is a web page on an e-commerce website that provides information on a specific product or item listing. Just below the details of the item on this page, the buyer can usually find recommendations for other relevant items. These are typically in the form of a series of modules or carousels, with each module containing a set of recommended items. The selection and ordering of these item recommendation modules are intended to increase discover-ability of relevant items and encourage greater user engagement, while simultaneously showcasing diversity of inventory and satisfying other business objectives. Item recommendation modules on the IDP are often curated and statically configured for all customers, ignoring opportunities for personalization. In this paper, we present a scalable end-to-end production system to optimize the personalized selection and ordering of item recommendation modules on the IDP in real-time by utilizing deep neural networks. Through extensive offline experimentation and online A/B testing, we show that our proposed system achieves significantly higher click-through and conversion rates compared to other existing methods. In our online A/B test, our framework improved click-through rate by 2.48\% and purchase-through rate by 7.34\% over a static configuration.
\end{abstract}

\ccsdesc[500]{Information systems~Learning to rank}
\ccsdesc[300]{Information systems~Novelty in information retrieval}
\ccsdesc[500]{Computing methodologies~Ranking}
\ccsdesc[500]{Applied computing~E-commerce infrastructure}

\keywords{A/B testing, page-level optimization}
\maketitle

\section{Introduction}\label{sec:intro}
\begin{figure*}[t]
 \centering
 \includegraphics[width=0.95\linewidth]{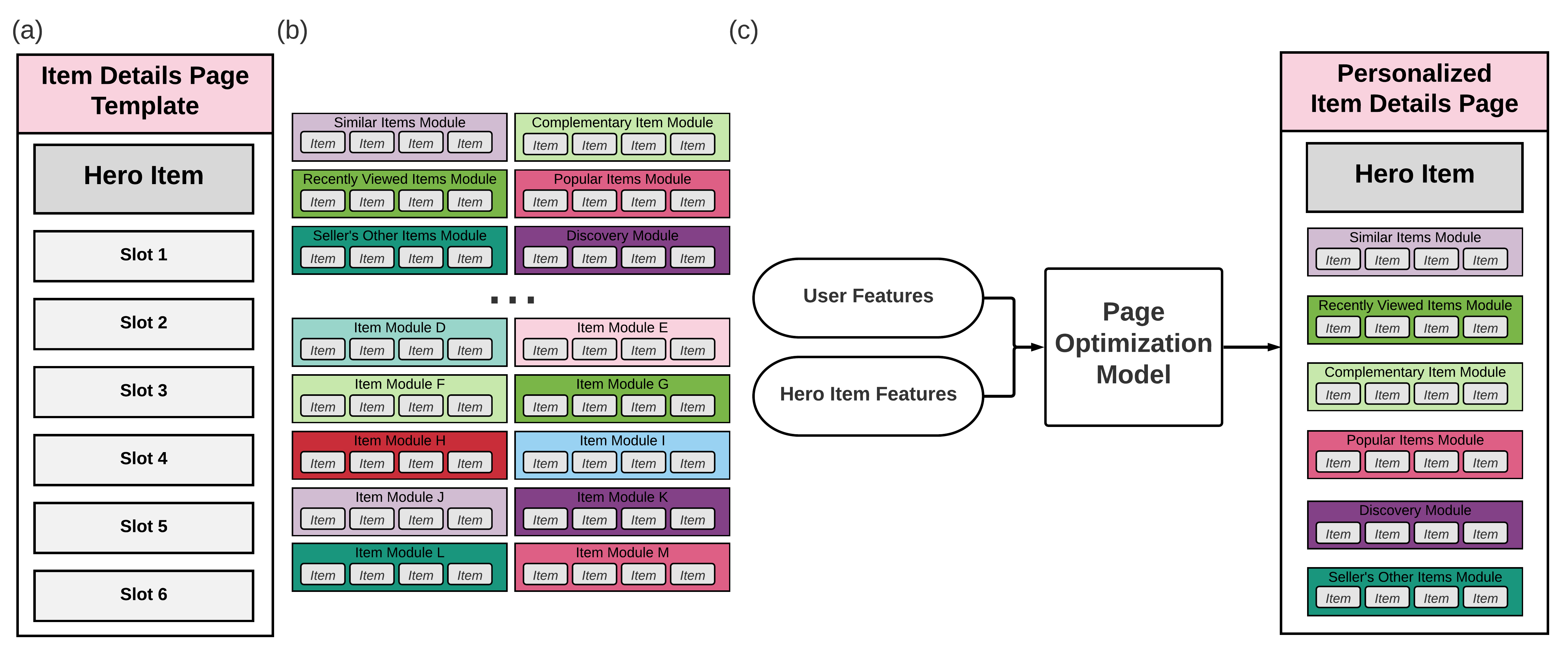}
 \caption{Overview of page-level optimization on item details page (IDP). (a) Typical layout of the IDP which shows the "hero" item and six recommendation modules. (b) Available module candidates. Note that modules with same color are in the same "module family" e.g., "Similar items module" and "Item Module J". (c) Given the page context including user features and "hero" item features, our page optimization model selects and orders available modules based on model scores.}
 \label{fig:po-overview}
\end{figure*}

E-commerce platforms such as Amazon, Alibaba, eBay, and Etsy all offer billions of live items for millions of users to consume. The item detail pages (IDP) that describe these items typically contain the item description and recommendations for other relevant items that the user may want to consider; this enables users to effectively browse relevant item inventory on the site (Figure~\ref{fig:po-overview}). One common paradigm of presenting such recommendations is to group them into carousels or modules with well defined themes. For example, a "similar items" module may recommend items similar to the "hero" item. A "complementary items" module may show recommendations that a user may find useful to buy along with the "hero" item.

These modules are typically powered by different recommendation algorithms, which may internally use different strategies for selecting and ordering items to present to the user. A lot of research in e-commerce recommendations have focused on improving these recommendation algorithms~\cite{Covington2016,He2014}; however, this is not the focus of this paper. In this paper, we study the problem of page-level optimization of recommendation modules on the e-commerce IDP. We assume we have a large selection of modules (powered by diverse recommendation algorithms) available to present to the user on the IDP, and we focus on the problem of optimizing the whole page presentation by selecting and ordering modules that best suit the current shopping intent of the user.

On a \textit{statically configured} IDP, the selection and ordering of the recommendation modules is fixed for all users and shopping intents, which can be sub-optimal. For example, a user who has recently browsed across different item categories such as shoes, clothes, and hats may actually benefit from seeing recommendation modules that offer items from all of these categories, even when they are looking at the IDP for a specific pair of shoes. Earlier studies~\cite{Wilhelm2018, Clarke2008} demonstrate that user engagement increases when users are presented with a diverse selection of content. In general, page-level optimization needs to satisfy multiple important goals~\cite{Ding:2019, mantha2020realtime, Hill:2017, Wang2016}. First, we must recommend relevant modules to users based on their previous shopping history. Second, we must recommend a diverse set of modules to encourage greater user engagement, and lastly, we must satisfy additional business goals or requirements (e.g., conversion rate, revenue).

Recently, deep neural networks have gained significant attention and success in the field of information retrieval. Several deep neural ranking models have been proposed that can convert heterogeneous and high dimensional data into lower dimensional embedding~\cite{Ai2019, Mottini2017}. Among those models, recurrent neural networks (RNNs), which have been widely used in natural language processing (NLP) tasks, can naturally encode the past information through the hidden states and affect the inference results at the current step. Consequently, RNNs are best suited in our scenario since the page presentation is a top-down structure (Figure~\ref{fig:po-overview}) and the user naturally scans from top to bottom. In this respect, we adapt RNN and modify the processing of context information such that the model can select the best page presentation for different users.

More specifically, we propose a two-stage RNN (TRNN) model that directly utilizes user, item context and various shopping signals (e.g., click, watch, purchase) by embedding them into a lower dimensional space and optimizing for both relevance (i.e., click through rate) and conversion (i.e., purchase through rate), while presenting a diverse set of recommendation modules. We address the combinatorial complexity at online inference time by greedily selecting the best module at each slot considering the modules selected for earlier slots. To avoid getting trapped at local optima, we apply beam search~\cite{Aubert2002} to ensure the final page presentation is diverse and has the best overall relevance to the user's shopping intent. 

Our contributions are three fold: First, we formulate the page-level optimization problem as a sequential ranking problem and model it using a RNN. Second, we demonstrate that our framework can accommodate heterogeneous feature types including item and near real-time user information to generate a personalized page presentation. Third, we show that our approach produces promising experimental results in both online and offline settings, and significant uplift in both relevance and conversion compared to the static page.

\section{Related Work}\label{sec:related}
A simple way to derive an ordering of recommendation modules to present on the page is to predict the relevance score of each module using a point-wise model and sort modules based on these scores. Relevance ranking has long been one of the core problems in information retrieval. Given a query, the retrieval system returns a list of ranked documents based on calculated relevance scores. There are numerous existing ranking techniques available for a wide range of applications including web pages~\cite{Agarwal2015}, videos~\cite{Covington2016} and advertisements~\cite{Chen2013, He2014, Graepel2010}. However, such approaches do not take into account the interactions between modules, which can lead to a sub-optimal user experience. Several studies~\cite{Devanur2016,Wilhelm2018} have shown that modules with similar content may decrease user satisfaction and hence the conversion of recommendations into sales.

At the other extreme, each unique whole page presentation may be considered as a candidate presentation to be scored, where the one with the highest score is presented to the user.  However, this requires evaluating a combinatorially large number of page presentations, which quickly becomes prohibitively expensive, especially due to page loading latency considerations when a large set of modules are involved. Recent works approximate the combinatorial space by considering pairwise interactions among modules~\cite{Hill:2017, Wang2016,Ding:2019}, and explore simple greedy procedures to learn the inter-dependencies~\cite{Hill:2017}. However, these models are not scalable when the given contextual features have high cardinality (e.g. categorical features), which restricts their usage for our application. Another line of research~\cite{Ding:2019} considers constrained objectives from multiple stakeholders for video recommendations and homepage optimization. The authors model page-level relevance by focusing on an online learning scenario using a simple linear model that models diversity via incremental diversity gain. However, the dependency between widgets is only captured by a set of diversity features, which may not accurately reflect the inter-dependency between modules. Their ranking model is also simplified to obtain tractable inference and learning algorithms. Thus, tractability comes at the expense of the model’s expressive power. 

Contextual Bandits provide a natural framework to learn from user engagement. The framework enables learning the required balance between helping users purchase items (i.e., exploitation) and promoting new items of interest (i.e., exploration). It has shown great promise in areas of news article recommendations and Ad recommendations~\cite{Lihong2010,Graepel2010}. This framework, however, requires a learning phase during which sub-optimal orderings may be shown to users, which may not be appropriate in some business scenarios. A more detrimental issue specifically for \textit{linear} contextual bandits is that the number of input features is limited and cannot be extended easily~\cite{Bendada2020}. Instead of utilizing the contextual bandit framework to optimize page presentation directly, we use one of the well-known contextual bandit methods --- Thompson sampling --- to collect our training and evaluation data (using a small amount of production traffic) with a balanced explore/exploit threshold, such that we can collect rich unbiased training data without significantly degrading our key business metrics.

More recently, neural architectures have been proposed to solve the ranking problem and have seen significant success when the ranking features have high dimensionality. Some recent works~\cite{Ai2019, Mottini2017} extract representations of the entire set of candidates for ranking, thereby taking into consideration all candidates when assigning a score for each candidate. All the information about interactions between candidates needs to be stored in the intermediate compact representation and extracted when scoring; this is impractical for a resource limited online production system that requires very low latency. To mitigate this problem, one can use a sequential decoding approach to assign candidates scores conditioned on previously chosen candidates, and then rerank the candidates \cite{bello2019}. However, this approach has two drawbacks: first, it requires a pre-ranked list and is sensitive and susceptible to its quality. Second, our problem scenario requires that we select and order modules from a large candidate set at the same time. To improve upon existing methods, we present an approach that explicitly encodes page context (including user information) and models the inter-dependency between modules using a RNN, allowing us to dynamically adapt page presentation based on the user's shopping journey and item information. 

\begin{table*}
  \caption{Notations and descriptions}
  \label{tab:notation}
  \begin{tabular}{cl}
    \toprule
    Notations & Descriptions \\
    \midrule
    $x$, $u$, $e$, $m$ & page context, user information, hero item information, module information \\
    $M^x_\mathcal{N}$, $M^x_K$ & set of $\mathcal{N}$ available modules under page context $x$, top-$K$ modules ranked by model \\
    $\phi$, $r^x_{m_i}$ & scoring function, scores of the $i$-th module given page context $x$ \\
    $L_i(\theta)$, $Q$ & loss function of the $i$-th objective with model parameter $\theta$, the number of objective \\
    $w_i$ & the weight of $i$-th objective for scalarization \\
    $s$, $t$ & slot $s$ on the page, timestamp $t$ \\
    $\mathcal{F}$ & set of available module families \\
    $\alpha, \beta$ & parameters of beta distribution \\
    $y_c \in \{0, 1\}$, $y_p \in \{0, 1\}$ & click and purchase labels \\
    $P$ & probability function \\
    $h_x$, $h_m$, $o^{LSTM}$  & hidden vector of page context, embedding of module, output hidden vector of LSTM cell \\
    $d_h$, $d_m$ & dimension of page context hidden vector, module embedding  \\
    $d_o$, $d_f$ & dimension of LSTM output vector, and module family embedding \\
  \bottomrule
\end{tabular}
\end{table*}

\section{Problem Formulation and Preliminary}\label{sec:formulation}
In order to help customers on the e-commerce platform find items of interest, we present a vertical ordering of $K$ item recommendation modules (carousels) on the IDP, with each module presenting a sequence of items (see Figure~\ref{fig:po-overview}). These modules may represent diverse types of recommendations (e.g similar to the "hero" item, vs. complementary to the "hero" item, vs. items inspired by the user's recent browsing history) and be powered by completely different recommendation algorithms. Optionally, the modules may be variants of the same algorithm, differing, for example, only in the item recall strategy and/or ranking model used. We assume that we have a large number (of the order of hundreds) of different modules available to present, and when a user visits the IDP, we need to select the best $K$ modules to present in vertical order to the user. Note that when we restrict the page presentation to be a ranked list without considering any interaction between modules, and assume that users are more satisfied if more relevant modules are placed on top, our page-level optimization application reduces to the traditional ranking problem~\cite{He2014,Lihong2010}. 

We assume further that every time a user lands on an IDP, we can obtain the page context ($x$) which includes user information ($u$) and "hero" item information ($e$) ("hero" item category, price, etc). Similarly, every module also possesses a module feature ($m$) that captures salient characteristics of the algorithm used to generate item recommendations for that module.  Given the page context ($x$), we denote the available set of modules as $M^{x}_{\mathcal{N}}$, and the module feature of the $i$-th module as $m_i$. The page optimization ranking model will take in page context ($x$), module feature ($m$), and rank the top-$K$ modules $M^{x}_{K}$ according to the score function $\phi$ (Table~\ref{tab:notation}). 

Note that the space of $M^{x}_{K}$ can be exponentially large if we consider the whole page presentation as an input instance. In this respect, our proposed scoring schema considers a point-wise scoring function that maps every candidate module (for a given page context) to a score. More specifically, without loss of generality, suppose we have $M^{x}_{\mathcal{N}}$ modules and have already selected a set of $M^x_{\mathcal{S}}$ modules for slots $1, 2, ..., s-1, s$, the scoring function at slot $s+1$ will compute the scores for the remaining $M^{x}_{\mathcal{N}} \setminus M^{x}_{\mathcal{S}}$ modules, and the module with the highest score will be selected for slot $s+1$. Since our ranker is a recurrent neural network, previously selected modules ($M^{x}_{\mathcal{S}}$) are implicitly captured through the hidden states. Hence, our module scoring function can be formulated as $r^x_{m_i} = \phi (m_i | M^{x}_{\mathcal{S}})$ where $i \in M^{x}_{\mathcal{N}} \setminus M^{x}_{\mathcal{S}}$. 

The optimal selection of modules must attempt to fulfill multiple objectives such as relevance to the user's shopping context and conversion into sales. Without loss of generality, assuming there are $Q$ objectives in our page optimization system, each objective can be mapped to a loss $L_i(\theta)$, where $i \in \{1, ..., Q\}$. Here $\theta$ denotes the model parameters. Optimizing for the $i$-th objective is equivalent to minimizing $L_i$. However these objectives are not completely correlated, and may lead to different page presentations if optimized for independently~\cite{Lin2019}. In other words, optimizing for these $Q$ objectives simultaneously is a non-trivial task since the optimal solution for one particular objective is usually sub-optimal for another. We therefore consider an aggregate loss where each loss term is weighted with $w_i$: $L(\theta) = \sum_i^{Q} w_i L_i(\theta)$, where $w_i \geq 0, \forall i \in \{1, ..., Q\}$. 

For the purposes of the work described in this paper, we consider three different losses. The first quantifies \textbf{relevance} and \textbf{interest} to the user by using  click signals ($L_c$). The second loss directly optimizes for user \textbf{conversion} by using purchase signals ($L_p$). The third loss ($L_u$) considers the user's \textbf{purchase intent} by means of signals such as adding an item to cart, watching an item, etc. These three losses are not completely independent of one another as the latter two can only happen when a click has occurred. However, a click does not necessarily guarantee any of the purchase intent signals or conversions. Since the objectives are dependent on each other, it is impractical to select an optimal set of weights manually. Instead, we learn the weights automatically by leveraging the multi-objective learning scheme proposed in~\cite{Cipolla2018} that treats objective weights as parameters to be learnt during training. We adopt this learning framework and compare its performance to our heuristically chosen weights in Section~\ref{sec:multitask}.

\subsection{Page presentation diversity}\label{diversity}
From previous studies, we observe that modules with similar content presented in consecutive positions can drastically affect the user experience and may lead to lower user engagement and conversion. In order to improve the diversity of page presentation, we first group similar modules into \textbf{module families} ($\mathcal{F}$) based on several criteria: item ranking algorithms, type of content, theme of the module, etc. At inference time, we impose the constraint that modules from the same module family may not be placed in consecutive positions; this greatly improves the perceived diversity of the page presentation. 

\section{Data Collection}\label{sec:data}
In this section, we describe the dataset we collected through Thompson sampling, and the session-based user intent attribution for augmenting the training and evaluation labels. We also present a slot position bias estimation framework to mitigate bias by applying inverse propensity scores (IPS) to the training objective~\cite{Wang2018}.

\subsection{Experimental dataset}
The page-level optimization task may be modeled as either a binary classification problem (each module on the page has either a positive or negative label) or as a ranking problem (e.g., using ranking loss). In either case, we require positive and negative samples as training instances. We define a module to be a positive instance if any item in that module has a user interaction (e.g. click, purchase, or purchase intent) signal. For a page, we typically present $K$ modules, and a page is considered as a positive page if any of the modules receives user interaction. To collect positive and negative samples, we gather implicit user interaction signals through offline logging, and downsample pages with no user interaction to prevent data imbalance. 

It is well known that it is expensive to collect data using pure exploration as it may negatively impact business metrics. As a consequence, it becomes necessary to utilize part of the production model's data, but this type of data is known for containing certain bias (e.g., slot position bias). We therefore gather data from two sources: exploration-oriented data obtained from Thompson sampling (see Section~\ref{TS}), and exploitation data from the production model. However, there are several obvious issues with this setup: the two data sources have (i) different data distributions, (ii) different degrees of bias, and (iii) data imbalance, since the volume of data collected from the production model is much greater than that from the Thompson sampling. In order to balance exploitation and exploration, we downsample the production data so it matches the exploration data in volume before training, and compensate for position bias using slot IPS (see Section~\ref{position}) in the loss computation during training; experimental results can be found in Section~\ref{sec:model_comp}. We also consistently apply diversity strategies while collecting data from both sources (see Section~\ref{TS}) to unify data distributions.

\subsection{Thompson sampling}\label{TS}
Thompson sampling~\cite{Russo2018} is a well-known contextual bandit algorithm to balance exploration and exploitation in decision making. In our setting, we use a batch update strategy to update the parameters for each module on a daily basis to reduce noise and production load. More specifically, Thompson sampling samples the score of module $i$ proportionally to the probability of generating clicks based on past $t$ observations: 
\begin{equation}
r^x_{m_i}(t + 1) \sim P(c^x_i(1:t))
\end{equation}
, where $r_{m_i}^{x}$ represents the reward for module $i$ given page context $x$, and $c^x_i(1:t)$ represents the past $t$ observations of module $i$. We then rank modules based on the reward (score) sampled from the distribution.

However, Thompson sampling is not suitable for high dimensional input spaces as the underlying model is generally constrained to be \textit{linear} for the posterior to be tractable and sampled efficiently. As our page context ($x$) has high cardinality, we instead consider a simplified formulation that only considers a few page context ($\hat{x}$) (e.g., "hero" item category), and samples directly from the posterior of a beta distribution: $r_{m_i}^{\hat{x}} \sim Beta(\alpha_i, \beta_i)$, where $\alpha_i, \beta_i$ are parameters of the beta distribution which are computed proportional to the number of clicks and total number of past observations of module $i$, respectively.

Since data from production is subject to diversity constraints (Section~\ref{diversity}), we apply diversification strategies in order to match the distribution of data from Thompson sampling to offline data from production. The first strategy scans the selected modules from top to bottom, and greedily swaps modules when two modules in consecutive positions belong to the same module family. The second strategy demotes other modules in a family by re-weighting rewards once a module from the family is selected; this ensures that the chances of selecting modules in the same family are small. We evaluate the impact of diversity measures through online A/B tests in Section~\ref{sec:online}.

\subsection{Feature engineering}
Our training data integrates different sources of features such as item information, module-level context, and user interaction history. In order to build the training dataset, we use ETL pipelines to enrich each training instance offline using these data sources.

Page context is primarily derived from device platform (e.g. desktop browser, mobile web, iOS/Android native app) and the "hero" item (title, category) presented on the IDP. Module-level context is obtained based on the type of recommendation algorithm used, the theme of the module, sources of item recall, and the module's historical performance (e.g. click through rate). Lastly, user features for personalization are obtained based on the user's search, viewing and purchase history, and their module affinity and interactions. We aggregate these user engagement signals at different levels of temporal granularity and incorporate them as features into our model. 

As a result, our training data contains diverse types of features: categorical features (e.g. item category), numeric features (e.g. module historical click through rate), and token based embedding (e.g. item title tokens). As neural networks are sensitive to the numeric values of the features, we ensure that all features are restricted to a small range of values. To achieve this, we inspect the feature distributions and perform normalization such as scaling, clipping and log transformations before feeding them into our deep model. 

\subsection{User purchase intent attribution}\label{sec:label}
We use user clicks and purchases as positive signals for model training and evaluation. These two signals express different shopping intentions. User clicks may indicate \textbf{interest} in the item, whereas purchase signals capture the user's holistic consideration of multiple aspects such as utility, price, shipping, appearance, and more. It is well-known that the number of all purchase events is orders of magnitude smaller than the number of all click events. We address this disparity by attributing purchase intent to click events that precede a purchase event based on user sessions, as described below:
\begin{enumerate}
    \item We order the user's engagement history, including both purchases and clicks, by timestamp.
    \item Events are split into sessions, with each session containing events with temporal gap not exceeding $T$ minutes.
    \item Within each session, if there is a purchase event, all preceding click events within the session are partially attributed towards the purchase. Each click signal is weighted by $\eta^{(t_p - t_c)/\epsilon T}$ for some constant $\epsilon > 0$, where $t_p$ and $t_c$ are timestamps of the purchase and click event respectively, and $\eta \in (0, 1)$ is a decay factor to differentiate the relative importance of click events.
\end{enumerate}

Instead of directly using purchase signals for optimizing conversion ($L_p$), we use the \textit{attributed} purchase intent signals as they encompass richer information about the user intent. In Section~\ref{sec:offline}, we use the attributed purchase intent signals to compute the attributed purchase intent loss ($L_{ap}$) while training our TRNN model, which serves as the baseline in most of our evaluations and comparisons in later sections.

\subsection{Slot position bias}\label{position}
As the modules on the IDP are rendered in sequential order, the probability of the user clicking on a module at different slot positions varies based on the position. To estimate the examination probability at each slot position, we periodically run a small amount of random traffic (less than 5\%) to shuffle modules on the IDP~\footnote{Note that we can also use Thompson sampling data to estimate slot position bias, but it could still possess certain bias due to the nature of exploitation.}. We use a position-based click model to estimate the slot position bias~\cite{Wang2018}. The click through rate (CTR) of a module $m$ at position $s$ is decomposed into the examination probability at that position $P_e(s)$ and the click probability once the module is examined $P(y_c = 1| m)$, i.e. $CTR(m, s) = P_e(s)P(y_c = 1| m)$ under the assumption of position-based click model. The actual $P_e(s)$ is difficult to obtain because $P(y_c = 1|m)$ is unknown. However, we can obtain the relative examination probability over position 1, $\frac{P_e(s)}{P_e(1)} = \frac{CTR(m, s)}{CTR(m, 1)}$, as the \emph{click propensity}. Similarly, we can compute the \emph{purchase propensity} using the same procedure. We can compute the IPS based on the computed click/purchase propensity and directly multiply the IPS with the training losses during training~\cite{Wang2018}. Evaluation results showing the impact of including the slot IPS are shown in Section~\ref{sec:model_comp}.

\section{Model}\label{sec:model}
In this section, we describe our model architecture for the page-level optimization problem. Assume we have $\mathcal{F}$ different module families, and each module family $i$ contains $F_i$ modules. Each module can be represented by a feature vector $m_{i, j} \in \mathbb{R}^n$, where $i, j$ represents the $j$-th module of the $i$-th family. Let $\pi \in \Pi$ denote a ranked sequence of modules, and $\pi_j \in \{m_{1, 1}, m_{1, 2}, ..., m_{F, F_F}\}$ denote the module selected at position $j$. For example, for $\pi = (m_{2, 3}, m_{3, 5}, m_{1, 4})$, at position $2$ ($\pi_2$), module 5 from the 3rd module family is selected.

\begin{figure}[t]
 \centering
 \includegraphics[width=1.0\linewidth]{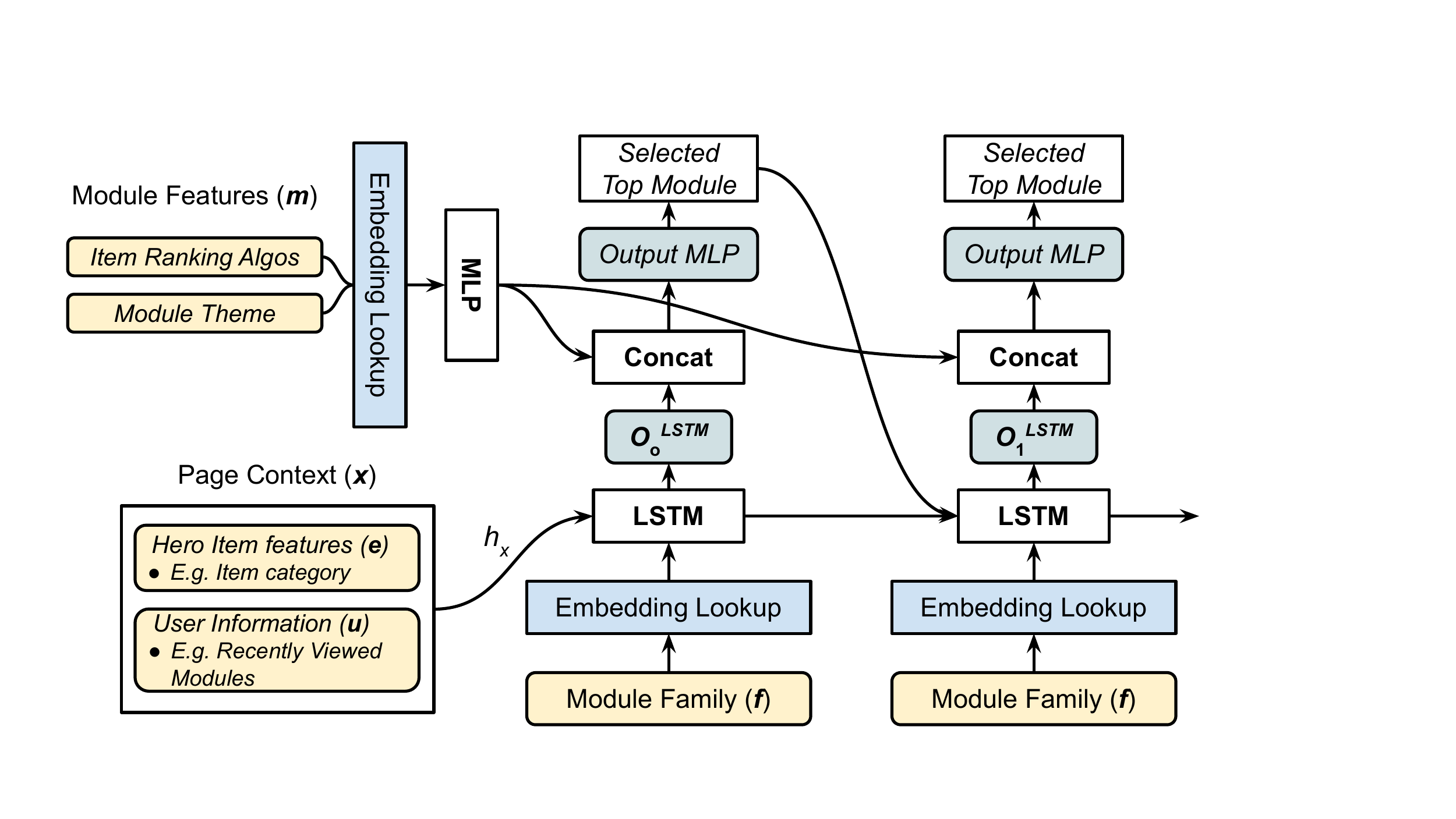}
 \caption{TRNN model architecture. The first stage processes the page context including "hero" item features and user information. The input to the LSTM cell is the available module family given page context ($x$). The output of the LSTM cell ($o_t^{LSTM}$) at step $t$ represents the latent information of all the module families, page context and the previously selected modules. In the second stage, we concatenate the module embeddings and then score them using the MLP output layer.}
 \label{fig:two-stage}
\end{figure}

For a general recurrent neural network (RNN), the output probability at position $j$ given the previous inputs can be expressed as a product of conditional probabilities according to the chain rule: 
\begin{equation}
\phi(\pi | x) = \prod_{j = 1} ^ K \phi(\pi_j | \pi_1, ..., \pi_{j - 1}, x)
\end{equation}
 where $x$ represents the page context. This expression is completely general and does not make any conditional independence assumptions. In the context of our problem, the scoring function $\phi(\pi_j| \pi_{<j}, x)$ models the probability of a module being placed at the $j$-th position in the page presentation given the modules selected for previous positions. For brevity, we have denoted the prefix permutation as $\pi_{< j} = (\pi_1, ..., \pi_{j - 1})$. This allows us to capture all higher order dependencies (e.g., diversity, similarity, and other interactions) between modules within the page. Note that the number of candidate modules can vary given different page contexts, but the number of module families remains constant across different settings. This inspires us to propose a two-stage network built upon the RNN (TRNN), as described below.

\subsection{Two-stage network architecture}
As shown in Figure~\ref{fig:two-stage}, our two-stage network is built upon a RNN that utilizes long short-term memory (LSTM) cells. In order to utilize the page context before the RNN decodes the sequence of the ranked modules, we encode page context ($x$) (which includes user information ($u$) and "hero" item information ($e$)) using a page-encoder network, which embeds the heterogeneous page context information into a $d_h$-dimension hidden vector $h_x$. $h_x$ serves as the hidden initialization of the LSTM cells at step 0, and we set the initialization of a cell state at step 0 to a $d_h$-dimensional zero vector. 

For the first stage, we embed module family and extract useful information through the LSTM cells. At each step $j \leq K$, the input ($f$) fed to the LSTM cell is a vector of module family features ($f \in \mathbb{R}^{F \times d_f}$), where $d_f$ is the dimension of module family embedding, and $F$ is the number of available module families in the given page context. The LSTM cell then outputs a $o_j^{LSTM} \in \mathbb{R}^{F \times d_o}$ vector, which is later concatenated with the output from the module embedding network (second stage). As previously discussed, the number of module families ($\mathcal{F}$) is much smaller than the the number of available modules. Hence the computation time for the LSTM input vector (module family embedding) will be much faster compared to naively feeding the module features directly.  

The module embedding ($h_m \in \mathbb{R}^{d_m}$) of module $i$ ($m_i$) is computed offline from the module network using a fixed set of module features (e.g., item ranking algorithms, module theme, etc). Hence, we only need to retrieve the pre-computed embedding from memory during online inferencing. Finally, the concatenation of the module embedding and the output of the LSTM ($o_j^{LSTM}$) is fed to an MLP and the output of the final softmax layer is used to compute the cross-entropy loss. To incorporate multiple user engagement signals, we tweak the architecture of the output MLP layer to account for multiple output signals. The aggregated multi-objective loss (Section~\ref{sec:formulation}) is trained using the learning strategy proposed in~\cite{Cipolla2018}.


\subsection{Beam search}\label{sec:beam_search}
During the inference phase, we apply beam search to find the optimal sequence of modules. At each step of inference, beam search maintains the $W$ ranked sequences with the highest scores. We define the scoring function ($\phi$) of a sequence up to slot $s$ for a beam $b$ $\pi_{< s+1}^b = (\pi^b_1, ..., \pi^b_s)$ as the probability of having at least one positive (purchased) module. Assuming the purchase event is independent of previously selected modules, we can compute $\phi(\pi^{b}_{< s+1})$ based on sequence $\pi^b_{< s}$:

\begin{equation}
\begin{split}
    \phi(\pi^b_{< s+1}) & = 1 - \prod_{i=1}^s (1 - P(y_p = 1 | \pi_i^b)) \\
    & = 1 - (1 - \phi(\pi^{b}_{<s} )) (1 - P(y_p = 1 | \pi_s^b))
\end{split}
\end{equation}

The final sequence will be selected as the one with the highest score in the last step. Note that greedy inference can be considered a special case where $W=1$. In our experiments, the bandwidth $W$ of beam search is fixed to 3 due to the production constraints. In Section~\ref{sec:online} we will compare the beam search strategy with greedy inference in the online setting.

\begin{figure*}[t]
 \centering
 \includegraphics[width=0.7\linewidth]{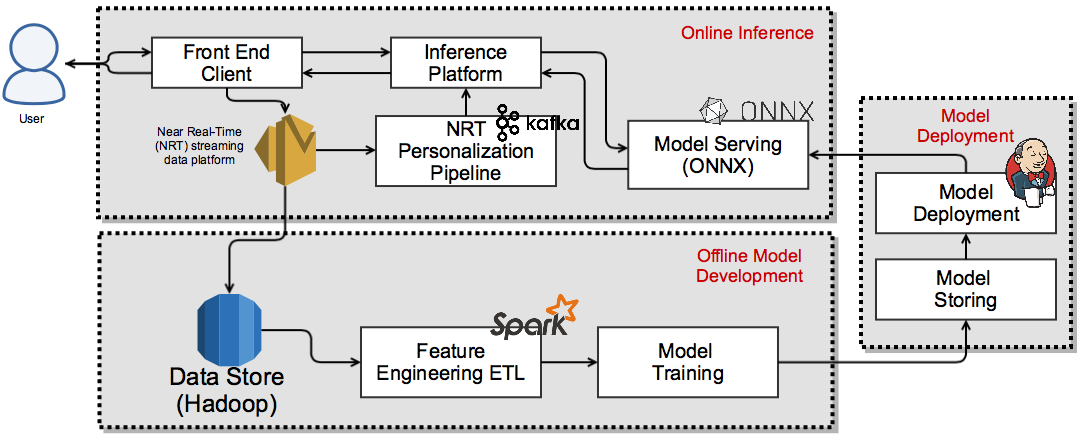}
 \caption{Production system architecture. Our production system consists of (1) online inference that coordinates the front-end request and execute back-end ONNX engine, (2) offline model development based on Spark feature engineering ETL and model training, and (3) model deployment and storing through Jenkins.}
 \label{fig:system}
\end{figure*}

\section{System Architecture}\label{sec:system}
In this section, we provide an overview of our production system (Figure~\ref{fig:system}), which consists of several components: online inference, offline model development, and deployment. We first describe our online inference platform and the near real-time (NRT) personalization pipeline, and then discuss offline feature engineering and model training. 

\subsection{Online inference platform and NRT personalization pipeline}
In order to generate optimal and personalized page presentations for the IDP with low latency, we built a scalable online inference platform for our production system. When a user visits the IDP, the "hero" item information and the user information are sent to the inference platform, which generates recommendations by executing the following steps: (i) Retrieve raw features from various feature stores including our NRT personalization engine that aggregates recent user activities and engagements to produce user-level features (e.g. user affinity to certain types of algorithms, recent purchases, interested categories, etc), (ii) Transform the raw features into feature vectors (e.g. convert "hero" item title and category into embeddings, normalize the feature values) (iii) Execute the TRNN ranking model with the beam search algorithm (Section~\ref{sec:beam_search}), which runs on an Open Neural Network Exchange (ONNX) back-end, and (iv) apply appropriate post-processing logic to satisfy business constraints (e.g. module diversity) and return the optimal page presentation to the user.

The presented module ranking and subsequent user engagement with the modules are recorded and tracked using a NRT streaming system. This stream of events is funneled into two downstream systems: a data warehouse for offline data analysis and model training, and our NRT personalization engine.

\begin{figure*}[t]
 \centering
 \includegraphics[width=0.8\linewidth]{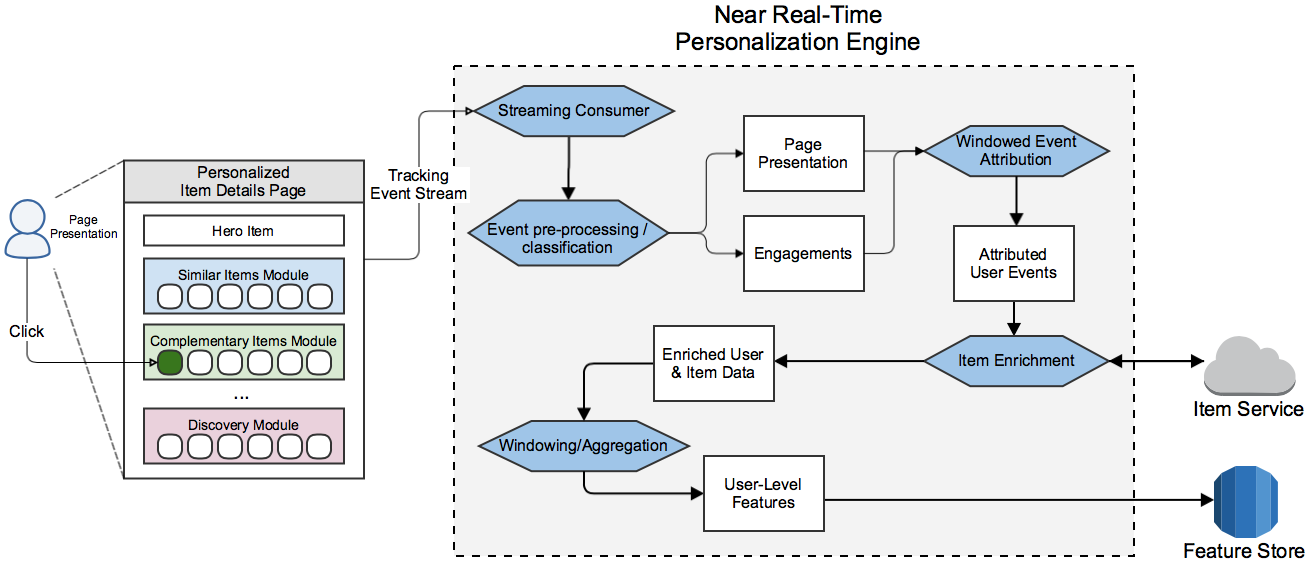}
 \caption{Near real-time personalization engine which consists of several stages that aggregate and storing the features.}
 \label{fig:personalization_engine}
\end{figure*}

The personalization engine (see Figure~\ref{fig:personalization_engine}) is a streaming system that listens to user engagement events. Each user event (e.g. page view, click or purchase) triggers a series of actions to recompute user-level features: (i) attribute the user engagement (e.g. clicks), which are often delayed responses, to the page presentation shown to the user to recover the original page context; (ii) query various databases to enrich details of the engaged items for insights into user interests (e.g. price/brand/category preference) and (iii) process and aggregate recent user activities and item interactions to compute user-level features. The updated user features are then written to a database, enabling instantaneous subsequent retrieval by the inference platform for module ranking.

\begin{table*}[!h]
  \caption{Model architecture comparisons. We consider the relative lifts ($\Delta$) to the XGBoost baseline. +$IPS$ means the model trained with inverse propensity score (IPS).}
  \label{tab:architecture}
  \begin{tabular}{ccccccccc}
    \toprule
     & $\Delta$AUC & $\Delta$F1 & $\Delta$NDCG & $\Delta$MRPR & $\Delta$CTR(DM) & $\Delta$PTR(DM) & $\Delta$CTR(DR) & $\Delta$PTR(DR) \\
    \midrule
    MLP & 5.09\% & 11.36\% & -35.82\% & -8.10\% & 2.24\% & 4.70\% & 5.80\% & 12.16\% \\
    MLP + $IPS$ & 5.20\% & 11.62\% & -35.07\% & -9.56\% & 1.77\% & 4.09\% & 5.21\% & 8.94\% \\
    TRNN & 7.32\% & 12.27\% & 27.81\% & 8.96\% & 11.00\% & 18.76\% & 11.65\% & 24.18\% \\
    TRNN + $IPS$ & 4.58\% & 5.61\% & 40.67\% & 16.39\% & 14.86\% & 24.12\% & 15.78\% & 29.20\% \\
  \bottomrule
\end{tabular}
\end{table*}

\subsection{Offline model development and deployment}
As discussed in the previous section, the logged page presentation and user engagements are stored in a data warehouse system, allowing us to retrain and redeploy the page optimization model using the latest data regularly in an offline setting. 

We utilize Apache Spark to build data pipelines for a variety of aforementioned data transformations: data sampling, slot position bias estimation, purchase intent attribution and feature engineering (see Section \ref{sec:data}). Next, we utilize GPU clusters to train our page optimization model and store the model candidates in the model repository. A model candidate is only signed off for production deployment after passing all validation checkpoints, including offline model evaluations, verification of ONNX model export, as well as load and performance tests for online inference latency. In addition, we analyze the predictive distribution of the model using a large pool of simulated pages and user contexts to detect any major shifts or deviations from the current production baseline.

\section{Offline Experiments}\label{sec:offline}
In this section, we describe offline evaluation of our proposed TRNN model and compare it with baselines for different scenarios. We consider two baselines: multilayer perceptron (MLP) and XGBoost~\cite{Chen2016}. The MLP model uses the same embedding method and a similar number of model parameters as the TRNN model for fair comparison. For the XGBoost model, we use most of the default parameter settings but set some parameters (e.g. maximum tree depth) to avoid overfitting to the training dataset. For the subsequent evaluation, all the models are trained with labels and weights that are attributed as described in Section~\ref{sec:label}.

\subsection{Evaluation metrics}
We experimented with multiple evaluation metrics to measure model performance using our offline validation dataset. Since our page optimization problem is similar to the ranking problem in the information retrieval setting, we consider several metrics commonly used for ranking problems such as Normalized Discounted Cumulative Gain (NDCG) and the Mean Reciprocal Purchase Rank (MRPR) which is similar to Mean Reciprocal Rank (MRR)~\cite{Chen2009}. We also evaluate how well the model computes the purchase probability of a module using F1 score. Given a page impression with at least one user engagement signal, we compute the recall, precision and report F1 score based on the binary classification scenario for the \textbf{purchase} signal. We then rerank the page impression based on the model output and compute the purchase NDCG and mean reciprocal purchase rank accordingly. 

\subsection{Unbiased offline evaluation}
In addition to the above-mentioned metrics, we also estimate the CTR (click through rate) and PTR (purchase through rate) to evaluate relevance and conversion, respectively. However, the click and purchase signals in the logged dataset could be biased since we only receive the feedback from the modules presented to the user. To overcome this issue, we adopt both Doubly Robust (DR) Estimator and Direct Method (DM) which evaluate both metrics in an unbiased manner~\cite{dudik2011doubly}.

\subsection{Model training}
We use the PyTorch deep learning framework to implement the core model. We chose the Adam optimizer~\cite{kingma2014} with a 0.001 learning rate. We collected around 1 million page presentations and split by 80\%/10\%/10\% as training, validation, and testing sets, respectively. 

We trained the TRNN model with 10 epochs over our data to reach convergence of the evaluation metrics. Model hyperparameters were selected considering production storage constraints and model performance on the validation set. The experimental results described below each used an average of 5 independent evaluations. 

\subsection{Evaluation results}
We consider three sets of experiments to demonstrate that our proposed TRNN model can improve performance compared with the two strong baselines. The embedding dimension for MLP and TRNN models are set to 50 and the number of linear layers are configured within the range of 1 to 3. 

The first experiment directly compares the intrinsic model performance with the consideration of slot position bias during training (see Section~\ref{position}). The second experiment considers the addition of personalized features, which we see can dramatically increase model performance. Finally, we consider optimizing for different user engagement signals during training, and show that considering click, purchase intents, and attributed purchase intents simultaneously can improve performance.

\subsubsection{Model comparisons}\label{sec:model_comp}
In this set of experiments, we first consider using the simplest set of features which excludes user information from the page context. During training, we also incorporate the slot IPS for the MLP and TRNN models to correct the bias induced by slot position. 

As shown in Table~\ref{tab:architecture}, the reported numbers are all relative to the XGBoost model, we can clearly see that the TRNN model outperforms the baseline with large margins in terms of all the evaluation metrics. However, the MLP model has poor biased ranking results compared to the XGBoost model, but obtain higher uplifts when evaluated using unbiased estimators. 

We also observe that training with slot IPS for TRNN models can increase the performance in terms of unbiased estimator which implies that we can correct the slot bias and rank relevant modules at higher slot positions. On the contrary, MLP does not benefit as much from considering slot position bias compared to the baseline MLP model. One potential reason is that MLP already uses slot position as one of its input features. 

\begin{table*}[!h]
  \caption{Relative lift from adding personalization features. The model without personalization features is used as the baseline for fair comparison. $+ pf$ corresponds to model with personalization features added. }
  \label{tab:personalization}
  \begin{tabular}{ccccccccc}
    \toprule
     & $\Delta$AUC & $\Delta$F1 & $\Delta$NDCG & $\Delta$MRPR & $\Delta$CTR(DM) & $\Delta$PTR(DM) & $\Delta$CTR(DR) & $\Delta$PTR(DR) \\
    \midrule
    XGBoost + $pf$ & -1.35\% & -4.83\% & 0.69\%  & 1.05\%  & 0.68\% & 1.04\% & 5.47\% & 0.11\% \\
    MLP + $pf$ & -0.13\% & 1.28\% & 1.83\% & 2.94\% & 0.76\% & 1.58\% & 0.71\% & 0.78\% \\
    TRNN + $pf$ & 0.16\% & 1.27\% & 1.82\% & 5.95\% & 1.93\% & 3.48\% & 7.69\% & 16.95\% \\
  \bottomrule
\end{tabular}
\end{table*}

\begin{table*}[!h]
  \caption{Impact of using different user engagement signals to train TRNN models. We consider user click ($L_c$), intent ($L_u$) and attributed purchase intent ($L_{ap}$) signals. We also consider $+ auto$ for learning the weights for each loss automatically. The baseline used is the model trained with only purchase intent signals ($L_{ap}$).}
  \label{tab:multi}
  \begin{tabular}{ccccccccc}
    \toprule
    & $\Delta$AUC & $\Delta$F1 & $\Delta$NDCG & $\Delta$MRPR & $\Delta$CTR(DM) & $\Delta$PTR(DM) & $\Delta$CTR(DR) & $\Delta$PTR(DR) \\
    \midrule
    TRNN ($L_c$) & 1.01\% & 0.56\% & 0.41\% & -4.54\% & -2.5\% & -4.36\% & -3.39\% & -5.26\% \\
    TRNN ($L_u$) & 1.13\% & 0.39\% & 2.03\% & -3.89\% & 0.69\% & 1.27\% & -1.17\% & -1.52\% \\
    TRNN ($L_c, L_u, L_{ap}$) & 0.04\% & 43.50\% & 1.09\% & 0.00\% & -1.44\% & -2.36\% & -1.56\% & -5.65\% \\
    TRNN ($L_c, L_u, L_{ap}$) $+ auto$ & 0.00\% & 43.13\% & -1.48\% & 2.40\% & 2.29\% & 3.86\% & -0.96\% & -0.96\% \\
  \bottomrule
\end{tabular}
\end{table*}

\subsubsection{Personalization features}
We next evaluate the impact of personalization features by comparing the performance of each model with and without personalization features. As can be seen in Table~\ref{tab:personalization}, all metrics see improvements, except for AUC and F1 score for XGBoost and MLP models. Overall it is clear from these results that user personalization features can truly add value to our page optimization models.

\subsubsection{Multiple user engagement signals}\label{sec:multitask}
We next evaluate the impact of training the TRNN model using multiple objectives. In this experiment, the baseline considers only the attributed purchase intent signal ($L_{ap}$). We consider the following training scenarios: (i) training solely on click signal and user purchase intent signals (e.g. watch signal), and (ii) multi-objective training using click, user purchase intent, and attributed purchase intent signals together with automated loss tuning proposed in~\cite{Cipolla2018}. 

From Table~\ref{tab:multi} we see that when only using click signal, most metrics show negative results. Also, the user purchase intent signal by itself is not sufficient to boost performance. Notably, when we naively weight each objective equally, we see drops in terms of unbiased evaluations, which indicates that relevance and conversion are not well balanced. Upon introducing objective weight learning, we observe uplift in CTR and PTR, which illustrates that the weights assigned to different objectives during training can truly affect  performance.

\section{Online Experiments}\label{sec:online}
In this section, we consider the online performance of TRNN models as measured using A/B tests. Models deployed to production are selected based on the offline evaluation results.

\begin{table}[!h]
  \caption{Online A/B test results for different models. All reported results are significant with p-value $<$ 0.05}
  \label{tab:online}
  \begin{tabular}{cccc}
    \toprule
     & TRNN & TRNN + $pf$ & TRNN ($L_c, L_u, L_{ap}$) + $pf$ \\
    \midrule
    $\Delta$CTR  & 2.40\% & 2.61\% & 2.48\% \\
    $\Delta$PTR & 4.40\% & 4.89\% & 7.34\% \\
  \bottomrule
\end{tabular}
\end{table}

As shown in Table~\ref{tab:online}, we can see that all models generate large uplift compared to the static page. Comparing the relative PTR lift between TRNN models, we can see that there are large uplifts when we consider both personalization features and multiple user signals trained together. 

We also compare performance of different randomization strategies when collecting training data. Thompson sampling generates a relative uplift compared to the full randomization scheme (1.69\% CTR, 1.74\% PTR). When the diversity strategy described in Section~\ref{TS} to encourage page diversity is applied, we observe 2.38\% CTR and 3.25\% PTR uplift, compared to pure Thompson sampling.

To demonstrate the usage of unbiased evaluation, we considered two variants of beam search during inference and deployed them in our production environment. One variant starts the beam search immediately at the first slot, whereas the other starts at a later slot (e.g., slot 3). From offline analysis, we found that beam search starting at a later slot has better performance (0.45\% CTR and 1.32\% PTR relative uplifts). In online A/B tests, we found similar behavior (1.40\% CTR and 2.19\% PTR uplift), confirming results observed offline. 

\section{Conclusion and Discussion}
In this paper, we introduced a two-stage RNN model that can leverage both contextual and personalized user signals to optimize page-level item recommendations. The proposed model allows us to explicitly model interaction effects between recommendation modules without needing to solve a combinatorial problem. Combined with greedy or beam search algorithms, this model also provides a highly cost-effective online inferencing solution for serving personalized page-level recommendations in near real-time.

By leveraging Thompson sampling in the data collection pipeline, we are able to collect data with less inherent biases for both model training and offline evaluation, while minimizing negative impact to key business metrics. Enabled by counterfactual examples collected through contextual bandit algorithms, we conducted unbiased offline experiments that show the efficacy of our proposed models over baselines for improving both ranking and business metrics.

We also demonstrated that such a system can be deployed at scale in production to serve millions of buyers on our e-commerce website responsively. Our online A/B tests using the deployed page optimization system showed a significant uplift to our key business metrics (e.g. user clicks and purchases) over a static page, indicating more relevant item recommendations and higher user satisfaction. Since our proposed framework can flexibly handle heterogeneous contextual and user features, it can easily be extended to other use cases beyond IDP with minimal modifications, e.g. optimizing recommendation modules on the homepage, and the checkout success page, in order to fully personalize the user's shopping journey.

To further improve our current system, we plan to explore the following: (i) While the beta-binomial Thompson sampling algorithm is an elegant solution to balance exploitation and exploration in data collection, our formulation in Section~\ref{TS} only allows categorical context and thus is not scalable to high-dimensional context space. We plan to address this limitation by using deep Bayesian bandit algorithms~\cite{Riquelme2018}. (ii) Recently, transformer architectures have been widely used in NLP and recommendation systems with significant gains in performance and training efficiency over RNNs. We plan to evaluate transformer-based page optimization models against RNNs offline and online.

\bibliographystyle{ACM-Reference-Format}
\bibliography{sample-base}
\end{document}